\documentclass[12pt]{article}

\usepackage[margin=1in]{geometry}
\usepackage{setspace}
\usepackage{amsmath, amssymb, amsfonts}  
\usepackage{graphicx}  
\usepackage{csquotes}  
\usepackage{titlesec}  
\usepackage{lipsum}  
\usepackage{dirtytalk} 

\titleformat{\section}{\normalfont\scshape\Large}{\thesection.}{1em}{}
\titleformat{\subsection}{\normalfont\scshape\large}{\thesubsection.}{1em}{}

\usepackage[authordate, doi=onlynd]{biblatex-chicago}
\bibliography{references}

\newtheorem{issue}{Issue}

\begin{document}

\title{Smart treaties: A path to binding agreements in international relations?}
\author{Niklas V. Lehmann}
\date{March 27, 2025}  


\maketitle

\begin{abstract}
Can we create binding agreements between nations? Recently, scholars have argued that blockchain technology enables us to do so.
Given that this could greatly affect the anarchical world order implied by state sovereignty, this remarkable claim is investigated thoroughly.
By focusing on the technical implementation of smart contracts between nations, this article finds that the potential to create binding agreements using blockchain technology is far more limited than recently suggested.
\end{abstract}




\section{Introduction}


There is no world government that can enforce agreements between nations. Consequently, international agreements must be self-enforcing (\cite{barrett_environment_2005}).
However, scholars have recently argued that binding agreements between nations can be realized with the aid of blockchain technology (\cite{reinsberg_fully-automated_2020}, \cite{rangeley_potential_2024}). These scholars argue that smart contracts, 
which are contracts stored as a computer program on a blockchain, allow for automated and impartial enforcement of international agreements.
Agreements between nations that are encoded in a smart contract on a blockchain shall be called \textit{smart treaties}. 
\textcite{reinsberg_fully-automated_2020} argues in his article that smart treaties can revolutionize the way nations interact and thus foster cooperation and stability.
Although \textcite{reinsberg_fully-automated_2020} investigates smart treaties in-depth, his article leaves the actual implementation of smart treaties open. Consequently, it is not fully clear whether smart treaties are actually technically feasible and can be implemented as imagined.
In order to find out whether smart treaties can foster cooperation among nations, this study analyses how smart treaties might actually be implemented. 
The result of this article is that the potential of smart treaties to create binding agreements is far more limited than \textcite{reinsberg_fully-automated_2020} suggests. 
 Most importantly, this article argues that it is hardly possible to achieve automated enforcement and that a trusted third party is necessary to arbitrate the smart treaty. Furthermore, since smart treaties need to operate with cryptocurrencies, they have many drawbacks compared to escrow agreements using fiat money. Therefore, it seems unlikely that smart treaties will be adopted and will have a great impact on international cooperation. 

The article proceeds as follows: Section 2 briefly provides the necessary background.  Section 3 investigates a hypothetical implementation of a smart treaty. In doing so, this paper also paves the way for future research. Section 4 discusses the results and concludes.

\section{Background} 

\textit{Smart contract} is a technical term for contracts created on blockchains.
A blockchain is a distributed history of transactions and account data that is managed by a set of network participants. Blockchains are predominantly used for the management of cryptocurrencies (\cite{nofer_blockchain_2017}).
A smart contract works similar to a ticket vending machine, which takes inputs (money) and creates outputs (tickets) based on its design (\cite{wilkens_smart_2019}). 
Blockchain users can create these digital entities, which can be likened to a legal person. Once a smart contract is created on a blockchain, it is an automaton and behaves as programmed. This is insofar interesting as it automates a contractual process, just like a ticket vending machine. This automatisation of contractual processes is interesting for international relations because of the potential to make agreements binding. 
An agreement that is binding is one where \say{its violation entails high penalties which deters the actor from breaking it} (\cite{peleg_introduction_2007}). International agreements are not binding because penalties cannot be enforced without disrespecting state sovereignty. However, \textcite{reinsberg_fully-automated_2020} argues that this state of anarchy can be altered with smart contracts because they execute automatically. The automatic execution would thus enforce penalties where currently no enforcement would be practical. 

The next section discusses technical limitations to this vision, but let us first ponder how remarkable the idea by \textcite{reinsberg_fully-automated_2020} is. 
 International relations would be completely changed if nations could make binding agreements, 
because it would be possible to create agreements that would currently be ineffective. The space of possible agreements would be strictly enhanced, allowing for more cooperation.
Given that lacking international cooperation is at the heart of many pressing problems, the importance of this finding cannot be understated. 
Thus, \textcite{reinsberg_fully-automated_2020} claims that \say{blockchain technology can promote international cooperation, under relatively benign assumptions [...] tapping unmet potential for cooperation [...]}.
The only other explicit mention of smart treaties in the current literature is by \textcite{rangeley_potential_2024}, who also argue that smart treaties could have \say{broad geopolitical consequences} because of \say{automatic enforcement}. \textcite{rangeley_potential_2024} mostly discuss potential applications, such as non-aggression treaties.

Since \textcite{reinsberg_fully-automated_2020} argues that we can and should establish smart treaties between nations, the next section investigates how such a smart treaty could be implemented and probes whether smart treaties can help enforce penalties.

\section{Implementation of smart treaties} 

We illustrate the real-world implementation of smart treaties with an example: We consider two hypothetical nations, which we unceremoniously call \say{Nation A} and \say{Nation B}. Without loss of generality we assume that Nation A and Nation B have been on the brink of warfare for some time. Both nations fear that a first strike advantage of either country may lead to war. Therefore, they attempt to create a non-aggression-treaty in which the aggressor must pay a penalty to the victim. Given that this transfer cannot be enforced by the victim, especially during a state of war, Nation A and B seek to create a smart treaty which will automatically transfer funds.
In the following sections, we'll refer to any factors currently hindering the use of smart treaties, as proposed, as an \textit{issue}.

\subsection{How is a treaty violation determined?}

Assume that Nation A claims an attack by Nation B, but Nation B denies this. How does the smart treaty determine whether an attack has actually occurred?
Trusted third parties, called \say{arbitrators}, are needed to resolve smart contracts (\cite{wilkens_smart_2019}).
A smart contract can only process the information that is on the blockchain. If the needed information is not part of the transaction and account history, it has to be provided to the smart contract by someone. 

\begin{issue}
A trusted third party is needed to arbitrate the contract. 
\end{issue}

 Since neither of the parties to the treaty can be trusted with signalling a violation, an additional account on the blockchain needs to get the authorization to signal a violation. This source of information needs to be trusted by all signatories to the treaty. Thus, if the source of information is not a machine itself, the treaty would be subject to the third party's judgment and there would be no \textit{automatic} enforcement. 

\textcite{reinsberg_fully-automated_2020} argues that the information can be delivered to the blockchain automatically 
and proposes two options. Nations are unlikely to seriously consider these for several reasons that I outline below: 

\begin{enumerate}
    \item \textcite{reinsberg_fully-automated_2020} proposes to use prediction markets to supply the information to the blockchain. Prediction markets are forums for trading contracts which yield payments conditional on the outcome of uncertain events (\cite{arrow_promise_2008}). Therefore, asset prices contain information about the likely outcome of events. For example, if the contract pays 1\$ if Nation A invaded Nation B, and the market price is close to 1\$, it is likely that Nation A will invade Nation B. 
    
Whilst prediction markets are  a great source of foresight, they still require that the objective truth is provided by a trusted third party to settle the final value of contracts (\cite{wolfers_prediction_2004}). In the example above, a trusted third party would have to judge whether Nation A did in fact invade Nation B, in order to determine that the contract can be exchanged for 1\$.
What \textcite{reinsberg_fully-automated_2020} actually refers to is known as a \textit{self-resolving} prediction market (\cite{ahlstrom-vij_self-resolving_2020}). 
Self-resolving prediction markets cannot be recommended for use in serious applications, least of all smart treaties.
 A self-resolving prediction market simply treats the asset price as an unverified truth. Once the price of the traded asset would reach a certain limit, this would signal an violation by a party to the treaty and the associated penalty would be paid out.\footnote{The existing studies on self-resolving prediction markets assume that the final asset price is determined at a random time which would not work for the smart treaties.}
There is no guarantee that traditional prediction markets will provide the truth, and they can be manipulated by individual traders (\cite{wolfers_prediction_2004}). Self-resolving prediction markets are even more prone to manipulation because they essentially constitute a bubble, i.e. the traded contracts have no fundamental value.
There is currently very limited evidence that self-resolving prediction markets work to produce the objective truth (\cite{lehmann_mechanisms_2024}).

Consequently, the treatment of prediction markets by \textcite{reinsberg_fully-automated_2020} is generally incorrect.
Since it seems extremely unlikely that nations will endow the power to rule over their treaties and funds to a market that is a bubble, self-resolving prediction markets are no substitute for a trusted third party. 

    \item The second proposition of \textcite{reinsberg_fully-automated_2020} is to use internet-connected devices to detect and broadcast information automatically to the blockchain.
    It seems unlikely that nations would set up such an agreement.
     Given that the devices would be internet-connected and effectively in charge of funds, great detail to security would be necessary. Equipment failure could cause a false alarm and accidentally trigger transfers.
    Furthermore, this would effectively substitute whoever runs and has access to the devices as a trusted third party.\footnote{We assume that it is technically possible to detect the information related to the treaty with certainty with an electronic device, which might be rather difficult and potentially expensive in practice.} 
    However, in technical terms, this method for achieving automatic enforcement is not infeasible. Rather, it is unlikely that electronic devices and information systems can be designed to be trustworthy for such important tasks. 
\end{enumerate}

\subsection{How are funds transferred between parties?}

How can the smart treaty transfer cash from the violating party to other parties to the agreement?
The smart treaty cannot directly access any reserves because (i) nations have authority over finances handled within their jurisdiction and (ii) the smart treaty can only transfer funds that are allocated to it, i.e. are \say{in its account}. The analogy of a ticket vending machine still works: The machine can only return funds that are already \textit{inside} the machine. Therefore, all required funds need to be staked by the nations beforehand. This also ensures that all parties have sufficient liquidity to cover penalties.
Consequently, the smart treaty is essentially an automatically managed escrow account. 
This brings with it a number of issues. 
Any currency exchanged on blockchains is by definition \textit{cryptocurrency}.\footnote{A reader has remarked that smart treaties could be implemented without any currency. Such an arrangement would be pointless and equivalent to a \say{traditional} treaty because it is the penalty, i.e. the transfer of currency which makes agreements binding.}
Nations need to obtain cryptocurrency and \say{freeze} it in the smart treaty. This is costly, but we will assume that the value of peace greatly outweighs those costs, so that Nation A and B are still better off choosing the smart treaty.\footnote{This is in fact a reason to not make the size of the penalty any bigger than it needs to be. However, since the penalty needs to be of a sufficient size so as to deter violating the smart treaty, there is a range of possible values, which we assume to be greater than zero. In practice, determining the right amount of funds would probably be difficult to negotiate as it would involve balancing costs and benefits for all nations involved.}

\begin{issue}
    Nations need to expect that the cryptocurrency retains most of its value.
\end{issue}

 If the cryptocurrency loses value, the smart treaty is at risk of suddenly becoming ineffective because the financial loss from violating the agreement may suddenly no longer outweigh the benefit of first-striking. 
In order to minimize risk to the smart treaty, nations would have to stake more money in a smart treaty than in a traditional escrow agreement using fiat money 
because cryptocurrencies have historically been far more volatile than fiat money (\cite{noauthor_annual_2018}).
This begs the question: How much money will be needed, and how can the nations obtain the needed amount of cryptocurrency? 

\subsection{Obtaining currency}

Most meaningful smart treaties would likely involve cryptocurrency worth billions of dollars. 
The best historical analogy to smart treaties are escrow agreements which were settled using an account with a trusted third nation. 
In 1981, Iran deposited 3.95 billion \$ in 2024 purchasing-power-adjusted value as part of the Algiers Accords (\cite{noauthor_escrow_1981}).
The Libyan government deposited 4.6 billion \$ in 2024 purchasing-power-adjusted value with a Swiss bank in 2003 to lift sanctions related to the Pan Am 103 bombings (\cite{katzman_us-libyan_2003}).

There are two ways to obtain cryptocurrency: Firstly, cryptocurrency can be purchased from others. Secondly, duties to ensure the smooth functioning of the blockchain, often referred to as \say{mining} or \say{minting}, can be carried out. These are rewarded with newly created currency. Let us investigate each in turn.

It will be difficult to obtain enough currency via mining, because there is not enough cryptocurrency created per unit of time.
%
As of 2024, approximately 450 Bitcoin are created per day, which represents a current value of around 30 million \$.\footnote{The Bitcoin blockchain currently expands by around 144 blocks of transactions per day. The reward for publishing a block currently stands at 3.125 Bitcoin and is steadily halving every couple of years.}
A nation that would want to obtain some of this newly created currency would have to invest significant computer resources to supporting the blockchain
and would then only capture a tiny fraction of this value.
Theoretically, it would be possible to invest more computational resources to support the blockchain and capture a larger share of the 30 million \$ per day. However, if a single actor were to supply a large share of computational resources, this would lower trust in the cryptocurrency, causing a decline in its value. This is because if one actor were to supply enough computational resources they could unilaterally re-write the recent transaction history (\cite{antonopoulos_mastering_2015}).\footnote{A single actor would need to process transactions faster than all others currently do, i.e. have more than 50\% of the total computational resources devoted to the blockchain. Thus, this is referred to as the 51\%-attack.} 
Mining Bitcoin to accumulate 4 billion \$ would take roughly three and a half years if a nation contributed one-tenth of the total computational power to the Bitcoin blockchain and it would take far longer on other blockchains because other blockchains have a smaller money supply and less currency is created per unit of time.

Instead, nations could purchase cryptocurrency from others. There is enough cryptocurrency in supply. As of 2024, \texttt{coinmarketcap.com} lists 13 cryptocurrencies with a money supply worth over 10 billion \$ at current market prices. The cryptocurrencies Tether and Ethereum each have a money supply valued at over 100 billion \$, and Bitcoins money supply is valued at over 1 trillion \$ at current market prices.
This currency is owned by individuals and can be bought directly from them or via exchanges such as the Chicago Mercantile Exchange or Binance.

\begin{issue}
 It may be difficult to obtain and later sell amounts of cryptocurrency large enough to make smart treaties effective, and doing so requires trust in cryptocurrency institutions or individuals.
\end{issue}

Although there is enough cryptocurrency, not all of it is for sale at current market prices, and thus it is not clear how much cryptocurrency a single actor could purchase. 
There is some evidence that it is possible to purchase a sufficient amount of Bitcoin, as the automotive company Tesla claims in its 2021 SEC report that they purchased 1.5 billion \$ worth of Bitcoin in a single quarter and have held around 2 billion \$ of Bitcoin at some point (\cite{noauthor_form10-k_2022}).
However, obtaining the needed cryptocurrency is not sufficient. 
Unless nations start to value cryptocurrency itself, the smart treaty is not effective if funds cannot be exchanged back into fiat money. Therefore, nations need to expect to be able to exchange large amounts of cryptocurrencies for fiat money in the future. 

Nations also need to trust the network of peers 
which runs and maintains any blockchain
and controls important decisions related to the blockchains functionality.
For example, in June 2016 cryptocurrency worth 50 million \$ was stolen from a smart contract on the Ethereum blockchain. The cause of this was a programming error in the smart contract.\footnote{This also showcases a minor issue: Depending on the complexity of the agreement, translating this agreement without error into code may be difficult.} The Ethereum community then decided to roll back the transaction history to undo this accident, which has \say{raised concerns about Ethereum’s
governance, because it violated the premise of decentralized applications running exactly as specified} (\cite{imine_ethereum_2018}).\footnote{Furthermore, the instances running and maintaining blockchains may be subject to strong incentives, e.g. the blockchain Tether is managed by a single company.}

\subsection{Account management}

In order to be able to obtain, hold and transfer currency, nations have to create accounts on the blockchain.
Every account on a blockchain is secured by a single private key. This is a random number which, similar to a password, is a unique identifier to access the account.  
If this private key of any party is lost or compromised, the account is compromised forever. It is impossible to \say{reset} a private key. 

\begin{issue}
 The smart treaty hinges on the private key security of all parties involved.
\end{issue}

Worse still, the smart treaty may lose its effect immediately
because the smart treaty would transfer funds to a compromised account in case of a violation.
This must not result in a breakdown of cooperation, but a new smart treaty will have to be arranged, using new accounts.
However, if the private key of the third party that arbitrates the contract is lost, then the treaty is completely ineffective because no violation can be signalled. Even worse would be a situation where the private key of the trusted third party is stolen by a hacker, who could then transfer funds to either nation. 
Adversarial intentions are aided by the fact that all details about the smart treaty would have to be public. 

\begin{issue}
 Any agreement will be fully public, including account details, funds staked and transfers.
\end{issue}

This is because blockchains by their very nature are publicly accessible transaction histories. Everyone with an internet connection will be able to read the smart treaty. 
However, the publicity may also be desirable as it may create mutual trust.

\subsection{Consequences of adopting cryptocurrencies at a nation state level}

Given that smart treaties must be public, they can become subject to public scrutiny.
As of 2024, cryptocurrencies are not considered legal or regulated in many countries (\cite{stolbov_what_2020}). This may have to change if these nations were to sign smart treaties. However, it is not completely inconceivable that nations change the legal status of cryptocurrency. El Salvador has made Bitcoin a legal tender (\cite{alvarez_are_2023}), but is an international outlier in this regard. 



\begin{issue}
    Governments may be unwilling to associate themselves with cryptocurrencies because this may negatively impact their reputation and reduce the effectiveness of monetary policy.
\end{issue}

Reasons for this could include the high energy consumption of blockchains,\footnote{Only blockchains that use proof-of-work to coordinate the addition of new blocks have a very high energy consumption. However, as of 2024, Ethereum is the only major blockchain which does not use proof-of-work.} which may be seen as anti-environmentalist (\cite{wendl_environmental_2023}), as well as their huge volatility and the association of cryptocurrency with crime and corruption (\cite{alnasaa_crypto-assets_2022}, \cite{noauthor_annual_2018}, \cite{foley_sex_2019}).
The adoption of Bitcoin by El Salvador has been met with international criticism and domestic reluctance (\cite{noauthor_imf_2022}, \cite{alvarez_are_2023}).


Furthermore, if nations were to adopt cryptocurrencies this could lead to an increased use of cryptocurrencies 
since cryptocurrencies might be seen as more legitimate and less risky.
For governments this would almost certainly be undesirable
because cryptocurrencies are not controlled by central banks, which could have more difficulty to reach their monetary policy goals 
(\cite{noauthor_annual_2018}).

\section{Discussions} 

This article details how smart treaties would be implemented in the real world, which highlights multiple issues. Based on these issues, this article concludes that smart treaties provide few comparative advantages over regular treaties. 
Essentially, smart treaties are an escrow on a blockchain. This comes with several issues that traditional escrows do not have. The main proclaimed advantage of smart treaties, automatic enforcement, is unlikely to be present. This result is in stark contrast with the article by \textcite{reinsberg_fully-automated_2020}, who assumes that automatic enforcement is possible.
\textcite{reinsberg_fully-automated_2020} observes: \say{Where Blockchain transactions require enforcement in the real world, the Blockchain is as ineffective as conventional international law [...]}. I agree with this assessment.
The main comparative advantage of smart treaties seems to be that they place funds outside of any nations jurisdiction. However, this comes at the cost of having to deal with cryptocurrencies. 

There are numerous limitations to smart treaties, some are not yet fully understood. It is not clear how much cryptocurrency a single nation could purchase and whether this amount would be sufficient. Additionally, the consequences of such a government endorsement of cryptocurrency also demand additional research. 
Furthermore, if smart treaties were to ever become reality, numerous challenges unrelated to their technical implementation will arise. These include novel legal and diplomatic challenges, such as how to administrate such a process. A related question would be why nations did not make more use of escrow agreements in the past. Last but not least, the theoretical welfare consequences of moving from a non-cooperative to a cooperative setting are not yet fully understood and deserve research in their own right.

\printbibliography

\section*{Funding declaration}

No funds were granted for this work.

\section*{Competing interests declaration}

There are no competing interests to disclose.

\end{document}